\documentclass[11pt,a4paper]{article}
\pdfoutput=1
\usepackage{jheppub}
\usepackage{amsmath,amssymb,bm}
\usepackage{graphicx}
\usepackage{subcaption}
\usepackage{hepunits}
\usepackage{braket}

\allowdisplaybreaks

\title{Thrust distribution in Higgs decays at the next-to-leading order and beyond}

\author[a]{Jun Gao,}
\emailAdd{jung49@sjtu.edu.cn}
\author[b]{Yinqiang Gong,}
\emailAdd{gongyq@pku.edu.cn}
\author[b]{Wan-Li Ju,}
\emailAdd{wanli\_ju@pku.edu.cn}
\author[b,c,d]{and Li Lin Yang}
\emailAdd{yanglilin@pku.edu.cn}

\affiliation[a]{INPAC, Shanghai Key Laboratory for Particle Physics and Cosmology, School of Physics and Astronomy, Shanghai Jiao Tong University, Shanghai 200240, China}
\affiliation[b]{School of Physics and State Key Laboratory of Nuclear Physics and Technology, Peking University, Beijing 100871, China}
\affiliation[c]{Collaborative Innovation Center of Quantum Matter, Beijing, China}
\affiliation[d]{Center for High Energy Physics, Peking University, Beijing 100871, China}

\abstract{
We present predictions for the thrust distribution in hadronic decays of the Higgs boson at the next-to-leading order and the approximate next-to-next-to-leading order.
The approximate NNLO corrections are derived from a factorization formula in the soft/collinear phase-space regions.
We find large corrections, especially for the gluon channel.
The scale variations at the lowest orders tend to underestimate the genuine higher order contributions.
The results of this paper is therefore necessary to control the perturbative uncertainties of the theoretical predictions.
We also discuss on possible improvements to our results, such as a soft-gluon resummation for the 2-jets limit, and an exact next-to-next-to-leading order calculation for the multi-jets region.
}

\begin{document}

\maketitle

\newpage

\section{Introduction}\label{sec:introduction}

The successful operation of the LHC and the ATLAS and CMS experiments have led to the discovery of the Higgs boson and completion of the standard model (SM) of particle physics~\cite{Aad:2012tfa, Chatrchyan:2012xdj}.
Precision test on properties of the Higgs boson including all its couplings with standard model particles becomes one primary task of particle physics at the high energy frontier.
Continuous operation of LHC has shown great success on refined study of the Higgs boson, for example, the recent discovery of the Higgs couplings with top quarks \cite{Sirunyan:2018hoz, Aaboud:2018urx} and bottom quarks \cite{Aaboud:2018zhk, Sirunyan:2018kst}.
On the other hand, the ability of the LHC or high luminosity (HL) LHC are limited on several aspects in the study of Higgs couplings. Due to the huge SM backgrounds, the accuracy of measurements on the Higgs signal strength cannot go below the order of 5\% \cite{CMS:2018qgz}. It is also very difficult to probe Yukawa couplings of the fermions of first two generations \cite{Gao:2013nga, Soreq:2016rae, Bishara:2016jga, Bodwin:2013gca, Kagan:2014ila, Zhou:2015wra, Koenig:2015pha, Perez:2015lra, Chisholm:2016fzg}, as well as possible invisible decay channels present in new physics models. Besides, the sensitivity to Higgs self-interactions are rather weak \cite{Goertz:2013kp, Sirunyan:2018two, CMS:2018ccd, Aaboud:2018ftw}.

To measure the Higgs properties with higher accuracy and to probe rare decay
modes of the Higgs boson, there have been a few proposals to build a future
lepton collider that can serve as a Higgs factory. These include ILC \cite{Behnke:2013xla}, CEPC \cite{CEPCStudyGroup:2018ghi}, CLIC \cite{Lebrun:2012hj} and FCC-$ee$ \cite{Gomez-Ceballos:2013zzn}.
At a lepton collider, e.g., the CEPC \cite{CEPCStudyGroup:2018ghi}, all decay channels of the Higgs boson can be measured in a model-independent way including possible invisible channels, and the total width can be reconstructed.
The projected precision on most Higgs couplings are at the percent level thanks to the clean environment \cite{An:2018dwb}. This is an order of magnitude improvement over the ability of the (HL-)LHC.

Precision experiments require equally precision theoretical predictions. To further scrutinize the SM and to look for possible new physics beyond, it is necessary to calculate higher-order corrections to the production and decay of the Higgs boson.
In this respect, there have been enormous advances in recent years. For example, the next-to-next-to-next-to-leading order (N$^3$LO) quantum chromodymamics (QCD) corrections to Higgs boson production via gluon fusion in the heavy top-quark limit \cite{Anastasiou:2015ema, Mistlberger:2018etf} and to Higgs boson production via vector boson fusion within the  structure function approach \cite{Dreyer:2016oyx}, the next-to-next-to-leading order (NNLO) corrections to Higgs boson production in association with a jet in the heavy top-quark limit \cite{Boughezal:2013uia, Chen:2014gva, Boughezal:2015dra, Boughezal:2015aha}, and the next-to-leading order (NLO) corrections to Higgs boson pair production with full top-quark mass dependence \cite{Borowka:2016ypz} have been known for some time.
The two-loop mixed QCD and electroweak corrections have also been calculated recently for the associated production of Higgs boson and a $Z$ boson at electron-positron colliders \cite{Gong:2016jys, Sun:2016bel, Chen:2018xau}.

In this work, we are concerned with the hadronic decays of the Higgs boson. Namely, the final-state consists hadrons initiated by quarks and gluons. This channel is particularly interesting for a future lepton collider, since it is rather difficult to be detected at hadron colliders.
This channel also provides a unique place to cleanly study non-perturbative aspects of QCD related to gluon jets.
Due to the hadronic nature of this channel, the cross sections receive sizeable QCD corrections. As a result, higher order calculations for various observables in this process are highly demanded.
The partial width for $H \to b\bar{b}$ is known up to the next-to-next-to-next-to-next-to-leading order (N$^4$LO), in the limit where the mass of the bottom quark is neglected \cite{Baikov:2005rw}. The partial width for $H \to gg$ has been calculated to the N$^3$LO in the heavy top-quark limit \cite{Baikov:2006ch}. We refer the readers to \cite{Denner:2011mq, Spira:2016ztx} for a complete list of relevant calculations.
At a more exclusive level, the fully differential cross sections for $H \to b\bar{b}$ have been calculated to NNLO in \cite{Anastasiou:2011qx, DelDuca:2015zqa} with massless $b$-quarks, and in \cite{Bernreuther:2018ynm} with massive $b$-quarks.

For hadronic decays, event shapes are a class of good observables. On one hand, they are infrared safe observables which can be theoretically calculated order-by-order in perturbation theory. On the other hand, they can be experimentally constructed from the hadron momenta without the need to specify a jet algorithm.
For Higgs boson decay, in particular, one of the authors has proposed to use event shapes such as thrust, hemisphere mass and $C$ parameter to distinguish final states induced by the $Hgg$ coupling and the $Hq\bar{q}$ coupling \cite{Gao:2016jcm}. This may help to probe possible new physics effects which modifies the light-quark Yukawa couplings.
It is also suggested in \cite{Li:2018qiy} to use jet energy profile to improve
the measurement of the $Hgg$ coupling.

In this work, we investigate the thrust distribution in the hadronic decays of the Higgs boson. Such decays can be induced by the effective coupling between the Higgs boson and gluons, and can also be induced by the Yukawa coupling between the Higgs boson and quarks. We discuss these couplings in Section~\ref{sec:formalism}. We then calculate the leading order (LO) and the NLO contributions to the thrust distribution in Section~\ref{sec:LONLO}. We find that the NLO corrections are rather large, and proceed to construct an approximate NNLO prediction in Section~\ref{sec:NNLOA}. We conclude in Section~\ref{sec:conclusion}.

\section{Formalism}
\label{sec:formalism}

In this work, we study the thrust distribution in hadronic decays of the Higgs boson. The thrust $T$ is defined by
\begin{align}
T \equiv \max_{\vec{n}} \frac{\sum_i | \vec{n} \cdot \vec{p}_i|}{\sum_i |\vec{p}_i|} \, ,
\end{align}
where $\vec{p}_i$ runs over the 3-momenta of the final state particles, and $\vec{n}$ is a 3-vector with unit norm. It is conventional to introduce the variable $\tau \equiv 1 - T$, which we will use extensively later. The limit $\tau \to 0$ corresponds to the final-state configuration of two back-to-back jets, and the limit $\tau \to 1/2$ corresponds to a nearly isotropic event.

Our calculations are based on the effective Lagrangian
\begin{align}
\mathcal{L}_{\text{eff}} &= \frac{\alpha_s(\mu) C_t(m_t,\mu)}{12 \pi v} O_g + \sum_q \frac{y_q(\mu)}{\sqrt{2}} O_q \nonumber
\\
&\equiv \frac{\alpha_s(\mu) C_t(m_t,\mu)}{12 \pi v} H G^{\mu\nu,a} G_{\mu\nu}^{a} + \sum_q \frac{y_q(\mu)}{\sqrt{2}} H \bar{\psi}_q \psi_q  , 
\label{eq:Leff}
\end{align} 
where $\mu$ is the renormalization scale, $v$ is the vacuum expectation value
of the Higgs field, $H$ represents the physical Higgs boson after electroweak symmetry breaking, 
and $G_{\mu\nu}^a$ is the field strength tensor of the gluon field. $\psi_q$ is the light quark
fields namely excluding top quark. The strong coupling $\alpha_s(\mu)$ and the Yukawa coupling $y_q(\mu)$ are renormalized in the $\overline{\text{MS}}$ scheme with $n_f = 5$ active flavors, i.e., with the top quark integrated out. The Wilson coefficient $C_t(m_t,\mu)$ comes from integrating out the top quark, whose perturbative expansion can be written as
\begin{align}
C_t(m_t,\mu) = 1 + \sum_{n=1}^{\infty} \left( \frac{\alpha_s(\mu)}{4\pi} \right)^n C_t^{(n)}(m_t,\mu) \, .
\end{align}
The coefficients $C_t^{(n)}(m_t,\mu)$ have been calculated up to N$^4$LO \cite{Inami:1982xt, Djouadi:1991tk, Chetyrkin:1997iv, Chetyrkin:1997un, Chetyrkin:2005ia, Schroder:2005hy, Baikov:2016tgj}. For our purpose, we need the results up to N$^3$LO, which are given by
\begin{align}
C_t(m_t,\mu) &= 1 + \frac{\alpha_s}{4\pi} \, 11 + \left( \frac{\alpha_s}{4\pi} \right)^2 \left[ L_t \left( 19 + \frac{16}{3} n_f \right) + \frac{2777}{18} - \frac{67}{6} n_f \right] \nonumber
\\
&+ \left( \frac{\alpha_s}{4\pi} \right)^3 \bigg[ L_t^2 \left( 209 + 46 n_f - \frac{32}{9} n_f^2 \right) + L_t \left( \frac{4834}{9} + \frac{2912}{27} n_f + \frac{77}{27} n_f^2 \right) \nonumber
\\
&\hspace{4em} -\frac{2761331}{648} + \frac{897943\zeta_3}{144}  + \left( \frac{58723}{324} - \frac{110779\zeta_3}{216} \right) n_f - \frac{6865}{486} n_f^2 \bigg] \, ,
\end{align}
where $L_t=\ln(\mu^2/m_t^2)$, and we have set explicitly the number of colors $N_c=3$ to shorten the expression.

We work in the limit of vanishing light quark masses, $m_q = 0$, while keeping the Yukawa
coupling $y_q$ non-zero. This treatment can be justified if new physics beyond the SM
leads to a different relation between $y_q$ and $m_q$ in the low energy effective theory.
The zero mass limit is a good approximation as long as $\tau\gg m_q^2/m_H^2$.

The massless (chiral) limit brings about a few simplifications to our calculation,
which we elaborate in the following. The first immediate effect is that the two 
operators in eq.~(\ref{eq:Leff}) do not interfere when computing squared-amplitudes. That is to say, for all final state $X$, the following interference term
\begin{equation} 
 \braket{ 0 | G^{\mu\nu,a} G_{\mu\nu}^{a} | X } \braket{ X | \bar{\psi}_q \psi_q | 0 }
\label{eq:interference}
\end{equation}
vanishes to all orders in the strong coupling $\alpha_s$. This can be easily seen since the QCD interactions preserve chirality in the massless limit, while the quark operator $O_q$ couples two quark fields with opposite chirality. Therefore, irrelevant of the final states, it is guaranteed that one of the two matrix elements in the above interference term vanishes.

The second simplification resides in the fact that the two operators in
eq.~(\ref{eq:Leff}) do not mix with each other under renormalization. To see
this, it is sufficient to show that the two matrix elements
$\braket{q_L \bar{q}_R | G^{\mu\nu,a} G_{\mu\nu}^{a} | 0}$ and $\braket{gg | \bar{\psi}_q \psi_q | 0}$ are zero. The vanishing of both matrix elements follows from the same argument on chirality in the above. As a result of this observation, the two coefficients $C_t(m_t,\mu)$ and $y_q(\mu)$ evolve independently under the renormalization group (RG). We have
\begin{align}
\frac{d}{d\ln\mu} C_t(m_t,\mu) = \gamma^t(\alpha_s(\mu)) \, C_t(m_t,\mu) \, , \quad \frac{d}{d\ln\mu} y_q(\mu) = \gamma^y(\alpha_s(\mu)) \, y_q(\mu) \, .
\end{align}
The explicit expressions for the anomalous dimensions $\gamma^t$ and $\gamma^y$ are known to third order in $\alpha_s$, and are collected in Appendix~\ref{app:FOingredients}.

Finally, we note that in the massless limit, the impact of integrating out the top quark on the quark operator is fully absorbed by the Yukawa coupling $y_q(\mu)$ defined in the 5-flavor scheme. This is slightly different from the massive case \cite{Chetyrkin:1997un}, where in addition to the flavor-decoupling in $y_q(\mu)$, there is an extra Wilson coefficient $C_2(m_t,\mu)$ coming into play. However, this coefficient arises purely from a similar effect as the operator mixing between $O_g$ and $O_q$. Since we have shown above that such a mixing is absent when $m_q = 0$, we can conclude that $C_2(m_t,\mu)$ equals unity to all orders in $\alpha_s$.

\begin{figure}[t!]
\centering
\includegraphics[width=0.3\textwidth]{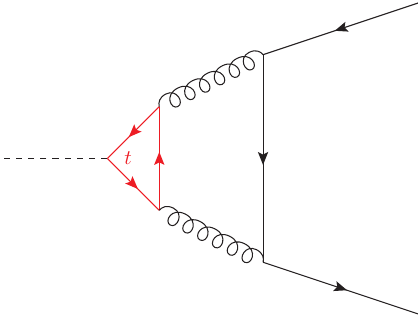}
\hspace{6em}
\includegraphics[width=0.3\textwidth]{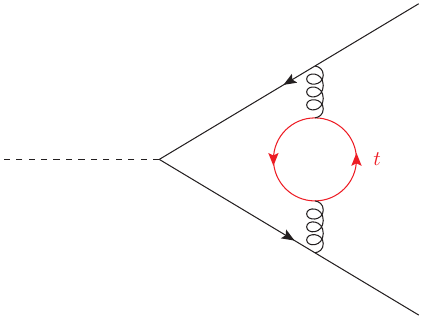}
\caption{\label{fig:HqqMatching}Representative top-quark loop contributions for the matching of the $Hq_L\bar{q}_R$ amplitude.}
\end{figure}

It is easy to demonstrate the above fact at the two-loop order (where the effect first appears). Consider the matching procedure for the $Hq_L\bar{q}_R$ amplitude. The matching coefficient comes from 3 contributions in the full theory with a closed top-quark loop: 1) diagrams where the external Higgs field is attached to the top-quark loop, e.g., the first diagram in fig.~\ref{fig:HqqMatching}; 2) diagrams where the Higgs filed is attached to the light quark propagator, e.g., the second diagram in fig.~\ref{fig:HqqMatching}; and 3) top-quark loop contributions to the renormalization of $y_q$ and $\psi_q$. The second and third contributions cancel each other if the renormalization constants for Yukawa coupling and quark field, $Z_y$ and $Z_\psi$, are chosen in the 5-flavor scheme. This cancellation is in fact the very definition of the ``5-flavor scheme'', which is obvious if we perform the matching with the external quarks on-shell and the Higgs momentum set to zero. As for the first contribution, it can be immediately seen that the first diagram in fig.~\ref{fig:HqqMatching} vanishes in the massless limit. The absence of the first contribution can be formally proven to all orders, since it is related to the on-shell matrix element $\braket{q_L\bar{q}_R | \bar{\psi}_t \psi_t | 0 }$.  Such an amplitude must have the form $F^\mu(p_1,p_2) \, \bar{u}(p_1) \gamma_\mu v(p_2)$ which is zero due to the equation-of-motion.

In summary, in the limit $m_q \to 0$, the hadronic decay of the Higgs boson can be classified at the parton level into two categories, induced by the gluon operator and the quark operator in eq.~(\ref{eq:Leff}), respectively. These two operators do not mix under renormalization. In the following, we will denote the partonic processes induced by the gluon operator as the $Hgg$ channel, and those induced by the quark operator as the $Hq\bar{q}$ channel. The names might sometimes be misleading, since the two channels can have the same final state particles. For example, the two operators can both induce the $H \to q\bar{q}g$ process. However, according to the discussions around eq.~(\ref{eq:interference}), these two amplitudes do not interfere with each other. As a result, from the computational point of view, we can strictly separate the $Hgg$ channel and the $Hq\bar{q}$ channel, and calculate higher order QCD corrections for them independently.

\section{The leading order and next-to-leading order results}
\label{sec:LONLO}

\begin{figure}[t!]
\centering
\includegraphics[width=0.3\textwidth]{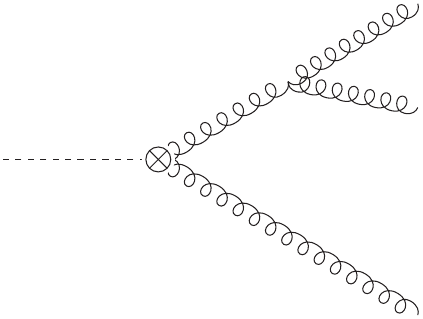}
\hspace{6em}
\includegraphics[width=0.3\textwidth]{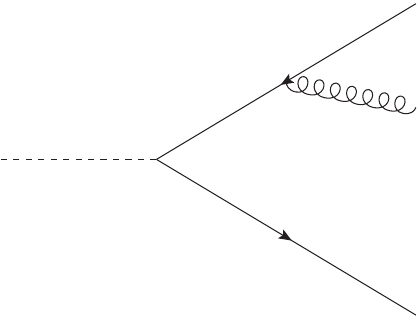}
\caption{\label{fig:LOdiagrams}Representative Feynman diagrams for the $Hgg$ channel (left) and the $Hq\bar{q}$ channel (right) for the thrust distribution at LO.}
\end{figure}

For the thrust distribution, at LO in $\alpha_s$, the $Hgg$ channel contains two partonic subprocesses $H \to ggg$ and $H \to q\bar{q}g$, while the $Hq\bar{q}$ channel has only one subprocess $H \to q\bar{q}g$. The representative Feynman diagrams are depicted in figure~\ref{fig:LOdiagrams}. The LO result for the $Hgg$ channel has been calculated in \cite{Mo:2017gzp}. We calculate the LO result for $Hq\bar q$ channel and also reproduce
the LO result for $Hgg$ channel. The expressions of normalized thrust distribution are given by
\begin{align}
\frac{1}{\Gamma^q_0} \frac{d\Gamma^q_{\text{LO}}}{d\tau} &= \frac{y_q^2(\mu)}{y_q^2(m_H)} \, \frac{\alpha_s(\mu)}{2\pi} \, C_F \, \frac{1}{\tau(\tau-1)} \left[ 3 (1-3\tau) (1-\tau)^2 - 2 \big( 2 - 3\tau + 3\tau^2 \big)
   \ln\frac{1-2\tau}{\tau} \right] , \nonumber
\\
\frac{1}{\Gamma^g_0} \frac{d\Gamma_{\text{LO}}^g}{d\tau} &= \frac{\alpha_s^2(\mu)}{\alpha_s^2(m_H)} \frac{\alpha_s(\mu)}{2\pi} \, \Bigg\{ C_A \, \frac{1}{3\tau(\tau-1)} \bigg[ (1-3\tau) (1-\tau) (11 - 24\tau + 15\tau^2) \nonumber
\\
&\hspace{15em} - 12 \big( 1 - \tau + \tau^2 \big)^2 \, \ln \frac{1-2\tau}{\tau} \bigg] \nonumber
\\
&+ T_F n_f \, \frac{2}{3 \tau} \left[ (1-3\tau) (2 - 15\tau + 15\tau^2) + 6 \tau \left( 1 -2 \tau + 2 \tau^2 \right) \, \ln\frac{1-2 \tau}{\tau} \right] \Bigg\} \, ,
\label{eq:LO}
\end{align}
where $\tau \in (0,1/3]$, $\mu$ is the renormalization scale, $\Gamma_{0}^{q} \equiv \Gamma_0^q(m_H)$ and $\Gamma_{0}^{g} \equiv \Gamma_0^g(m_H)$ are LO partial decay widths at the scale $\mu=m_H$, with
decay width at a scale of Higgs mass, with
\begin{equation}
\Gamma_{0}^{q}(\mu) = \frac{y_q^2(\mu) \, m_H \, C_A}{16 \pi} \, , \quad \Gamma_{0}^{g}(\mu) = \frac{\alpha_s^2(\mu) \, m_H^3}{72\pi^3 v^2} \, .
\end{equation}

\begin{figure}[t!]
\centering
\begin{tabular}{cc}
\includegraphics[width=0.3\textwidth]{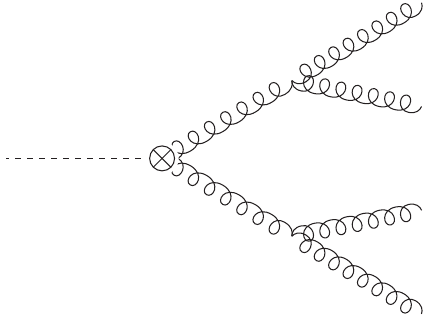}
&\qquad
\includegraphics[width=0.3\textwidth]{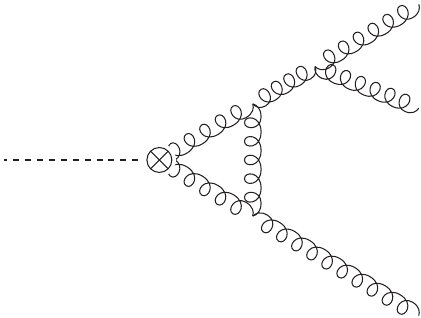}
\\[2ex]
\includegraphics[width=0.3\textwidth]{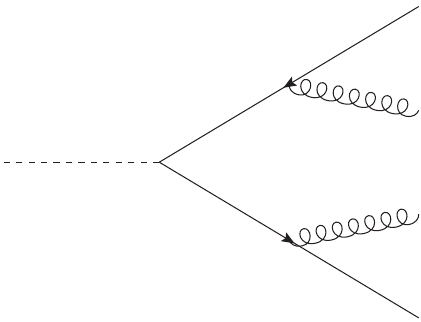}
&\qquad
\includegraphics[width=0.3\textwidth]{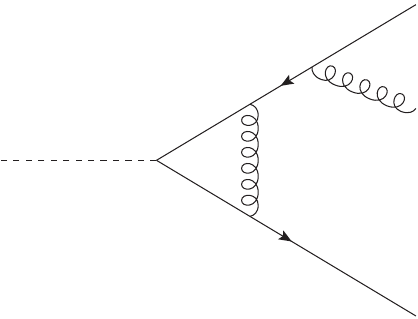}
\end{tabular}
\caption{\label{fig:NLOdiagrams}Representative Feynman diagrams for the $Hgg$ channel (upper) and the $Hq\bar{q}$ channel (lower) at NLO.}
\end{figure}

The NLO corrections to the thrust distribution involve both virtual gluon exchanges and real gluon emissions. The representative Feynman diagrams are show in figure~\ref{fig:NLOdiagrams}. The virtual diagrams contain ultraviolet (UV) divergences which are removed by renormalization of the couplings $\alpha_s$ and $y_q$. The renormalization constants are given by
\begin{align}
Z_{\alpha_s} = 1 - \frac{\alpha_s}{4\pi} e^{-\epsilon\gamma_E} (4\pi)^{\epsilon} \frac{\beta_0}{\epsilon} + \mathcal{O}(\alpha_s^2) \, , \quad Z_y = 1 - \frac{\alpha_s}{4\pi} e^{-\epsilon\gamma_E} (4\pi)^{\epsilon} \frac{\gamma^y_0}{2\epsilon} + \mathcal{O}(\alpha_s^2) \, ,
\end{align}
where $\beta_0$ and $\gamma^y_0$ are given in Appendix~\ref{app:FOingredients}; $\epsilon=(4-d)/2$ is the dimensional regulator; and $\gamma_E$ is the Euler constant. After renormalization, both real and virtual corrections are separately infrared (IR) divergent, while their sum is finite. In order to implement the cancellation in a Monte-Carlo generator, we adopt the dipole-subtraction method \cite{Catani:1996vz}. This amounts to introducing an auxiliary function $d\Gamma_A$ which has the same singular behaviors in the soft and/or collinear limits. The sum of the virtual and real corrections then be written in the form
\begin{align}
\Gamma^i_{\text{V+R}} = \int_{n+1} d\Gamma^i_{\text{real}} + \int_{n} d\Gamma^i_{\text{virt}} = \int_{n+1} \left( d\Gamma^i_{\text{real}} - d\Gamma^i_{A} \right) + \int_{n} \left( d\Gamma^i_{\text{virt}} + \int_1 d\Gamma^i_{A}\right) ,
\end{align}
where the integral symbol with subscript $n$ denotes an $n$-body phase-space integration, and $i=q,g$ represent the $Hq\bar{q}$ and $Hgg$ channels, respectively. The two terms in the above formula are both finite, and the integration can be performed numerically. For the $Hgg$ channel, there is an extra contribution from the $C_t$ coefficient at NLO. Combining everything, we have the NLO decay rates as
\begin{align}
\Gamma^q_{\text{NLO}} &= \Gamma^q_{\text{LO}} + \Gamma^q_{\text{V+R}} \, , \nonumber
\\
\Gamma^g_{\text{NLO}} &= \left( 1 + \frac{\alpha_s}{2\pi} C_t^{(1)}(m_t,\mu) \right) \Gamma^g_{\text{LO}} + \Gamma^g_{\text{V+R}} \, .
\label{eq:GammaNLO}
\end{align}

Based on the above formulas, we construct an in-house Fortran program to compute the differential decay rates. We use the real-emission matrix elements from OpenLoops \cite{Cascioli:2011va} and the one-loop matrix elements from Refs.~\cite{Schmidt:1997wr, DelDuca:2015zqa}. The Monte-Carlo integrations are performed with the Cuba library \cite{Hahn:2004fe}. For the input parameters, we use $\alpha_s(m_Z)=0.1181$, $m_H={125.09}$~GeV and $m_t={173.5}$~GeV. 

\begin{figure}[t!]
\centering
\includegraphics[width=0.45\textwidth]{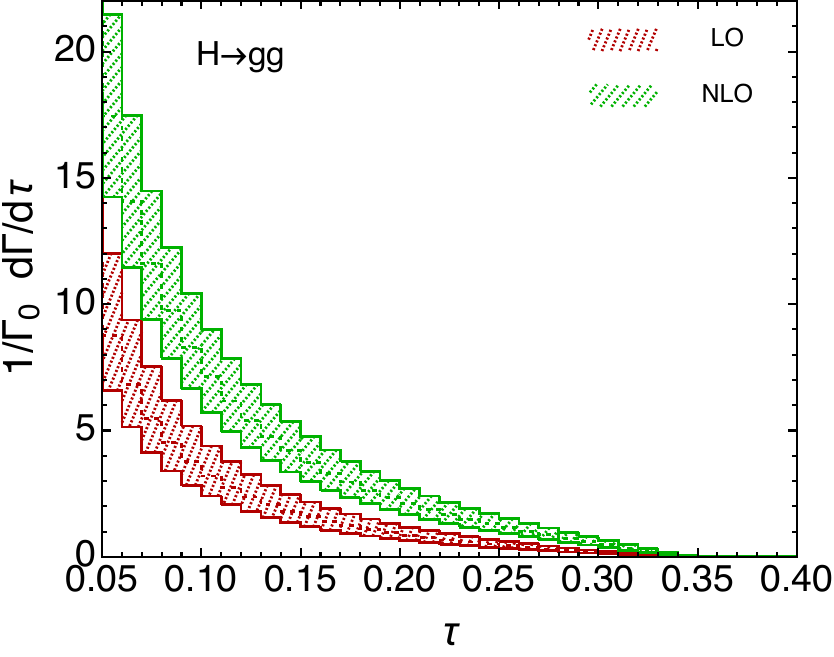}
\quad
\includegraphics[width=0.437531\textwidth]{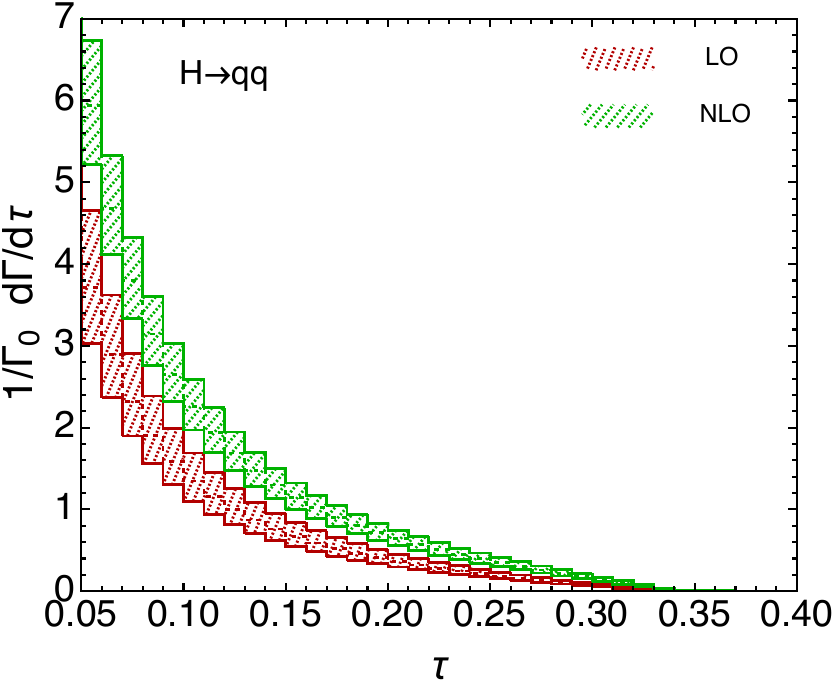} 
\caption{\label{fig:LONLO}Thrust distributions at LO and NLO in the $Hgg$ (left plot) and $Hq\bar{q}$ (right plot) channels.}  
\end{figure}

\begin{figure}[t!]
\centering
\includegraphics[width=0.45\textwidth]{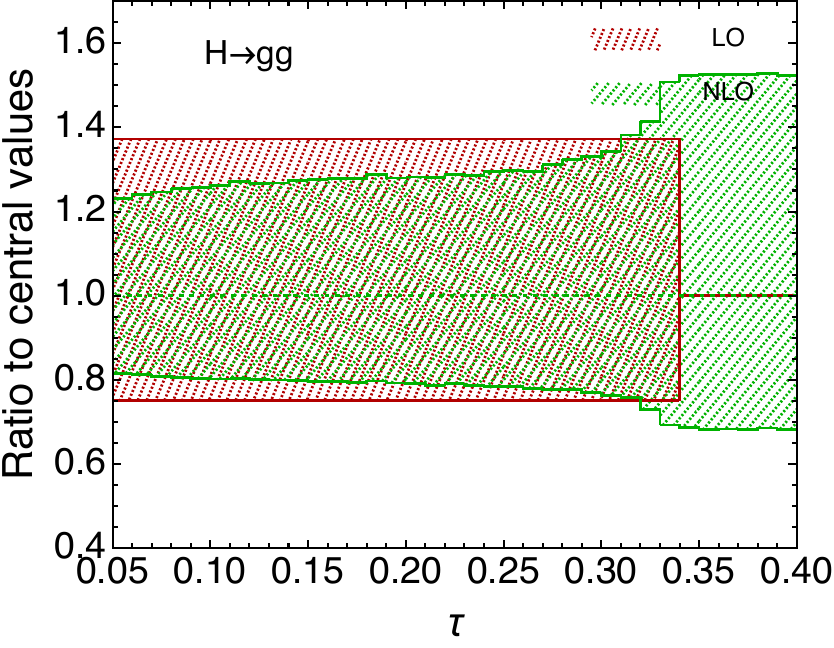}
\quad
\includegraphics[width=0.45\textwidth]{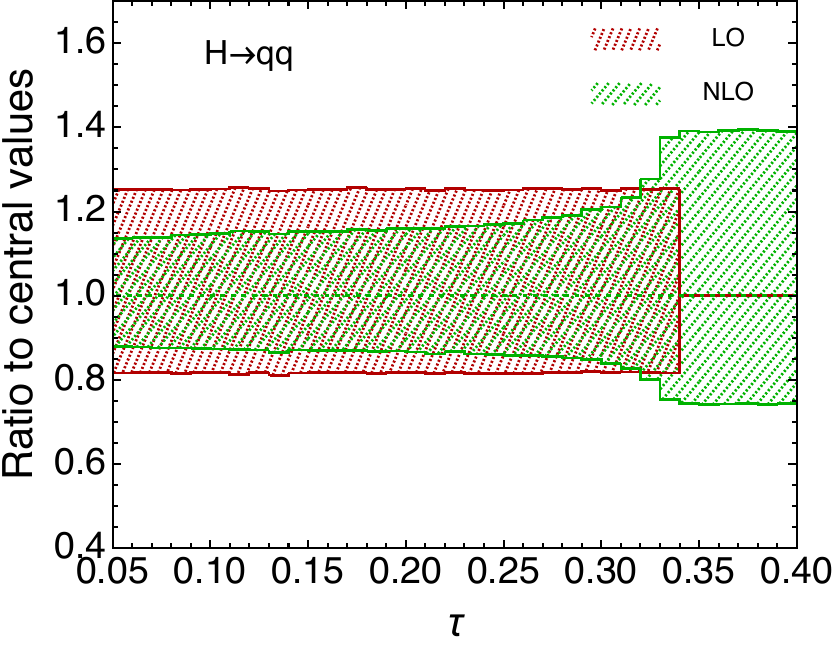}
\caption{\label{fig:LONLOscale}The ratios of the LO and NLO differential cross sections to their central values.}  
\end{figure}

In figure~\ref{fig:LONLO}, we show the LO and NLO thrust distributions in the $Hgg$ and $Hq\bar{q}$ channels, respectively. The error bands reflect the variations of the results when the renormalization scale $\mu$ is varied up and down by a factor of 2 from the nominal choice of $m_H$. Note that the LO distributions approach zero when $\tau \to 1/3$, due to phase space constraints. At NLO, with an additional parton emitted, the region $1/3 < \tau < (1-1/\sqrt{3})$ opens up. From this figure, one can see that the NLO corrections are rather large for both channels, indicating the bad convergence of the perturbative series. Especially for the $Hgg$ channel, the NLO differential cross section is twice the LO one at $\tau \sim 0.05$. The correction is even more pronounced for larger $\tau$. We also find that the scale uncertainties of the LO results do not overlap with the NLO ones. This indicates that the scale variation of the LO differential cross sections underestimate the theoretical uncertainties. We also show in figure~\ref{fig:LONLOscale} differential cross sections normalized to their central values. At LO the scale variations arise entirely
from running of the couplings and show no dependence on kinematics. The scale variations are reduced at NLO for $\tau$ below the kinematic endpoint at LO.  

To summarize, our NLO calculation reveals a few unsatisfactory features which make us believe that even higher order corrections are phenomenologically important. To obtain the full NNLO thrust distribution for, e.g., the $Hq\bar{q}$ channel, one needs to calculate, among others, the two-loop virtual corrections to the $H \to q\bar{q}g$ process, the one-loop virtual corrections to the $H \to q\bar{q}gg$ process, and the tree-level $H \to q\bar{q}ggg$ process. One also needs to combine these contributions, either analytically or numerically, in order to cancel the infrared divergences. Before get into such an involved computation, it is useful to estimate the size of the NNLO corrections. The rest of this paper will be devoted to the calculation of thrust distributions at approximate NNLO based on a factorization formula in small-$\tau$ limit. The factorization formula can also be used to resum large logarithms appearing in small-$\tau$ region, where the perturbative expansion is doomed to fail. This will be left to a future work in preparation.

\begin{figure}[t!]
\centering
\includegraphics[width=0.45\textwidth]{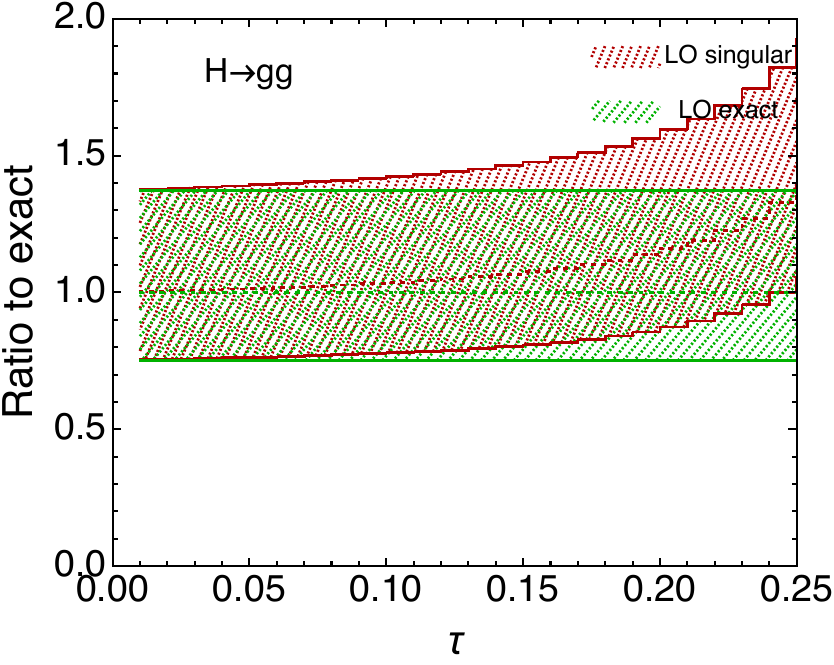}
\includegraphics[width=0.45\textwidth]{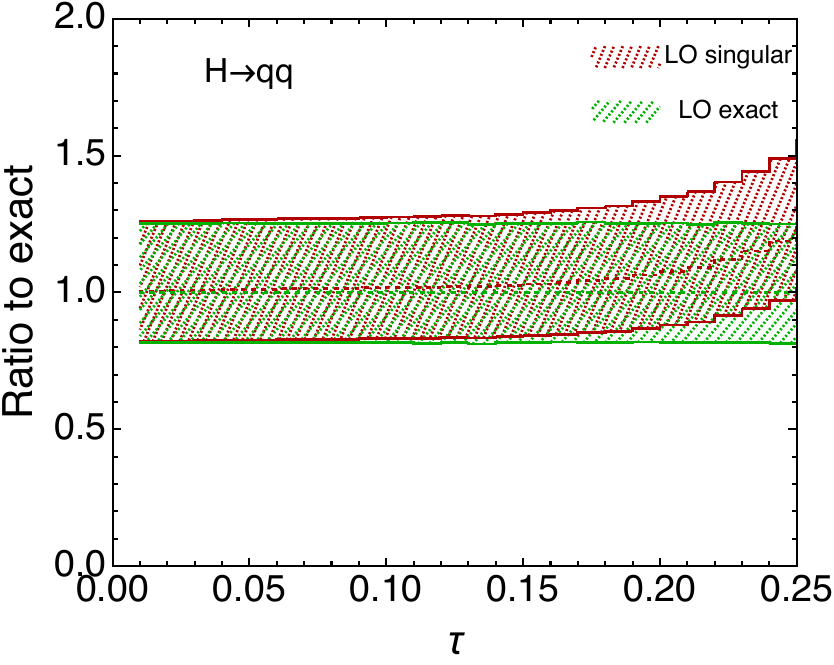}
\caption{\label{fig:LOsing}Comparison between the exact results and the singular terms at LO. }
\end{figure}

The factorization formula deals with singular terms of the form $\ln^n\tau/\tau$ in the thrust distributions. Before going into the NNLO corrections, we can extract such singular terms in the LO results from eq.~(\ref{eq:LO}). The results are given by
\begin{align}
\frac{1}{\Gamma^q_0} \frac{d\Gamma^q_{\text{LO,sing}}}{d\tau} &= \frac{y_q^2(\mu)}{y_q^2(m_H)} \, \frac{\alpha_s(\mu)}{2\pi} \, C_F \, \frac{1}{\tau} \big( - 4\ln\tau - 3 \big) \, , \nonumber
\\
\frac{1}{\Gamma^g_0} \frac{d\Gamma_{\text{LO,sing}}^g}{d\tau} &= \frac{\alpha_s^2(\mu)}{\alpha_s^2(m_H)} \frac{\alpha_s(\mu)}{2\pi} \left[ C_A \, \frac{1}{3\tau} \big( - 12\ln\tau - 11 \big) + T_F n_f \, \frac{4}{3 \tau} \right] .
\label{eq:LOsing}
\end{align}
In figure~\ref{fig:LOsing}, we compare numerically the singular terms at LO against the exact results by plotting their ratios.
From there one can see the singular terms dominate at small-$\tau$ region. They remain as the leading contributions up to $\tau\sim 0.25$,
where the non-singular terms contribute about 30\% and 20\% for $Hgg$ and $Hq\bar q$ respectively.

\section{Factorization at small $\tau$ and approximate NNLO}
\label{sec:NNLOA}

In this section, we briefly introduce the factorization formula at small $\tau$, and use it to derive an approximate NNLO formula for the thrust distribution. In the $\tau \to 0$ limit, the final state hadrons form two nearly back-to-back jets in the rest frame of the Higgs boson. In this reference frame, it is convenient to choose two light-like vectors $n=(1,0,0,1)$ and $\bar{n}=(1,0,0,-1)$ to represent the directions of the two jets. The momenta of the two jets are then labeled by $p_n$ and $p_{\bar{n}}$.
The factorization formula can be obtained using the language of soft-collinear effective theory (SCET) \cite{Bauer:2000ew, Bauer:2000yr, Bauer:2001yt, Beneke:2002ph, Beneke:2002ni, Becher:2014oda}, following the derivations for the $e^+e^- \to q\bar{q}$ process \cite{Schwartz:2007ib, Becher:2008cf, Bauer:2008dt}. The factorized form is given by
\begin{multline}
\frac{d\Gamma^i}{d\tau} = \Gamma_0^i(\mu) \, |C_t^i(m_t,\mu)|^2 \, |C_S^i(m_H,\mu) |^2 \int dp_n^2 \, dp_{\bar{n}}^2 \, dk \, \delta \bigg( \tau - \frac{p_n^2+p_{\bar{n}}^2}{m_H^2} - \frac{k}{m_H} \bigg)
\\
\times J_n^i(p_n^2,\mu) \, J_{\bar{n}}^i(p_{\bar{n}}^2,\mu) \, S^i(k,\mu) \, ,
\label{eq:fac}
\end{multline}
where $i=q,g$ denote the $Hq\bar{q}$ and $Hgg$ channels, respectively. We have defined $C_t^g(m_t,\mu) \equiv C_t(m_t,\mu)$ and $C_t^q(m_t,\mu) \equiv 1$, corresponding to the matching coefficients discussed in section~\ref{sec:formalism}.

The formula eq.~(\ref{eq:fac}) involves several ingredients, which we introduce in the following. The hard Wilson coefficients $C_S^i(m_H,\mu)$ comes from integrating out the hard fluctuations at the scale $\mu \sim m_H$. They are defined as the matching coefficient from the full theory eq.~(\ref{eq:Leff}) to SCET. They can be obtained from the $Hq\bar{q}$ and $Hgg$ form factors, which are know up to the 3-loop order \cite{Harlander:2003ai, Gehrmann:2005pd, Moch:2005tm, Gehrmann:2010ue, Gehrmann:2014vha}. From these results, the Wilson coefficients $C_S^q$ and $C_S^g$ can be extracted up to the next-to-next-to-next-to-leading order (N$^3$LO). The jet functions $J_n^i(p_n^2,\mu)$ and $J_{\bar{n}}^i(p_{\bar{n}}^2,\mu)$ describe collinear emissions along the directions of the two jets. The typical jet masses are given by $p_n^2 \sim p_{\bar{n}}^2 \sim \tau m_H^2$. Both the quark jet function and the gluon jet function have been calculated to the N$^3$LO \cite{Becher:2006qw, Becher:2010pd, Bruser:2018rad, Banerjee:2018ozf}. The soft functions $S^i(k,\mu)$, on the other hand, describe soft emissions with typical momenta $k \sim \tau m_H$. The quark soft function has been known analytically up to the NNLO \cite{Schwartz:2007ib, Fleming:2007xt, Kelley:2011ng, Monni:2011gb}. For our purpose, we also need the scale-dependent part of the N$^3$LO soft function, which can be obtained through its RG equation. Note that the scale-independent part of the N$^3$LO soft function was also extracted numerically, albeit with large uncertainty \cite{Bruser:2018rad}. Up to the N$^3$LO, the gluon soft function can be obtained from the quark one by a Casimir scaling $C_A/C_F$. The explicit expressions for the above ingredients are collected in Appendix~\ref{app:LSingredients}.

Given the factorization formula eq.~(\ref{eq:fac}), it is straightforward to obtain the leading singular terms for the thrust distribution by expanding the formula in terms of $\alpha_s(\mu)$. Up to the NNLO, the singular part of the thrust distribution can be formally written as
\begin{equation}
\frac{d\Gamma^i_{\text{sing}}}{d\tau} = \Gamma_0^i(\mu) \left[ \frac{\alpha_s(\mu)}{4\pi}\Delta_i^{(1)}(\tau,\mu) + \left( \frac{\alpha_s(\mu)}{4\pi} \right)^2 \Delta_i^{(2)}(\tau,\mu) + \left( \frac{\alpha_s(\mu)}{4\pi} \right)^3 \Delta_i^{(3)}(\tau,\mu) \right] ,
\label{eq:singular}
\end{equation}
with $i=q,g$. The explicit expressions of the coefficients $\Delta_i^{(n)}(\tau,\mu)$ can be found in Appendix~\ref{app:singular}.

\begin{figure}[t!]
  \centering
  \includegraphics[width=0.45\textwidth]{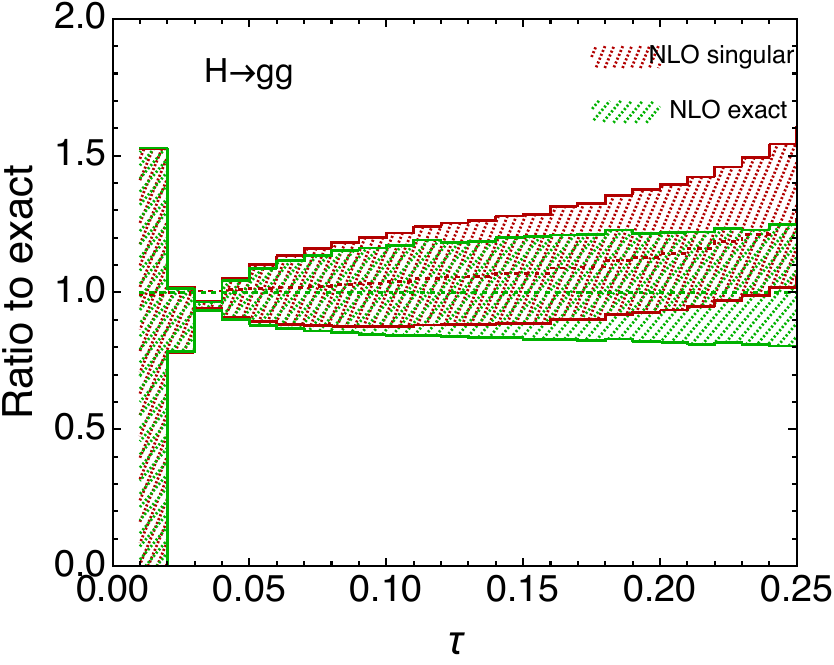}
	\includegraphics[width=0.45\textwidth]{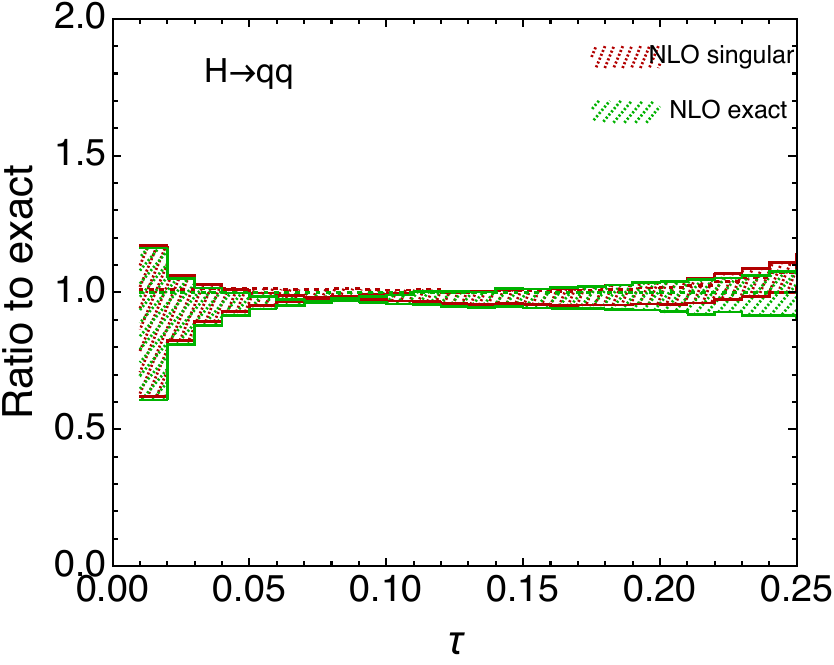}
\caption{\label{fig:NLOsing}Comparison between the exact results and the singular terms at NLO.}
\end{figure}

With the above formula, we can now perform a comparison similar to fig.~\ref{fig:LOsing} for the NLO corrections. This is shown in fig.~\ref{fig:NLOsing}. Again we see that the $\Delta_i^{(2)}$ term serves as a very good approximation of the exact NLO correction up to $\tau \sim 0.2$. This leads us to believe that the $\Delta_i^{(3)}$ term should also provide a good description of the NNLO correction in this region. Therefore, we define our Approximate-NNLO (NNLO$_{\text{A}}$) thrust distribution as
\begin{equation}
\frac{d\Gamma^i_{\text{NNLO,A}}}{d\tau} = \frac{d\Gamma^i_{\text{NLO}}}{d\tau} + \Gamma_0^i(\mu) \left( \frac{\alpha_s(\mu)}{4\pi} \right)^3 \Delta_i^{(3)}(\tau,\mu) \, .
\end{equation}
Namely, we add the NNLO singular contribution from $\Delta_i^{(3)}$ to the exact NLO result calculated in the previous section.

\begin{figure}[t!]
\centering
\includegraphics[width=0.45\textwidth]{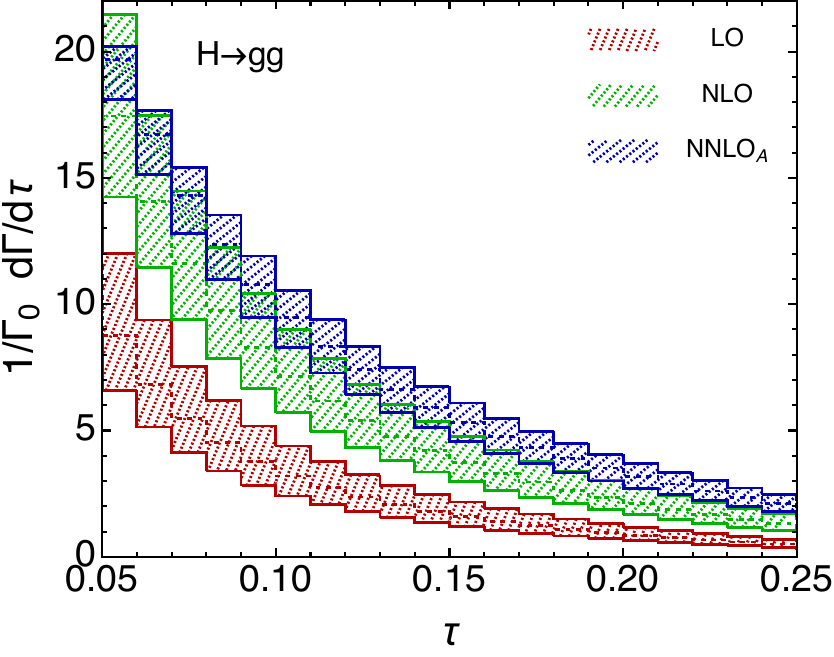}
\includegraphics[width=0.438545\textwidth]{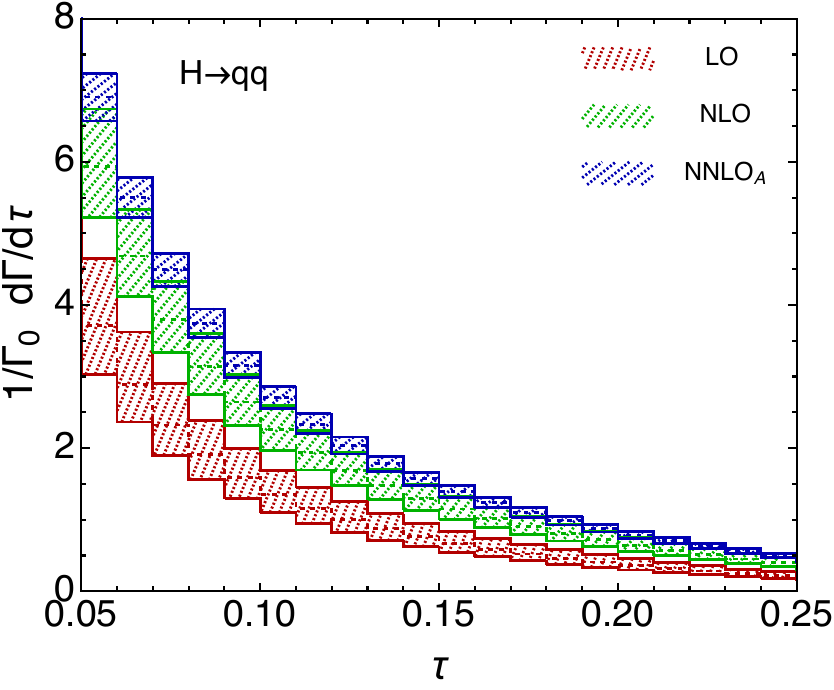}
\\
\includegraphics[width=0.437581\textwidth]{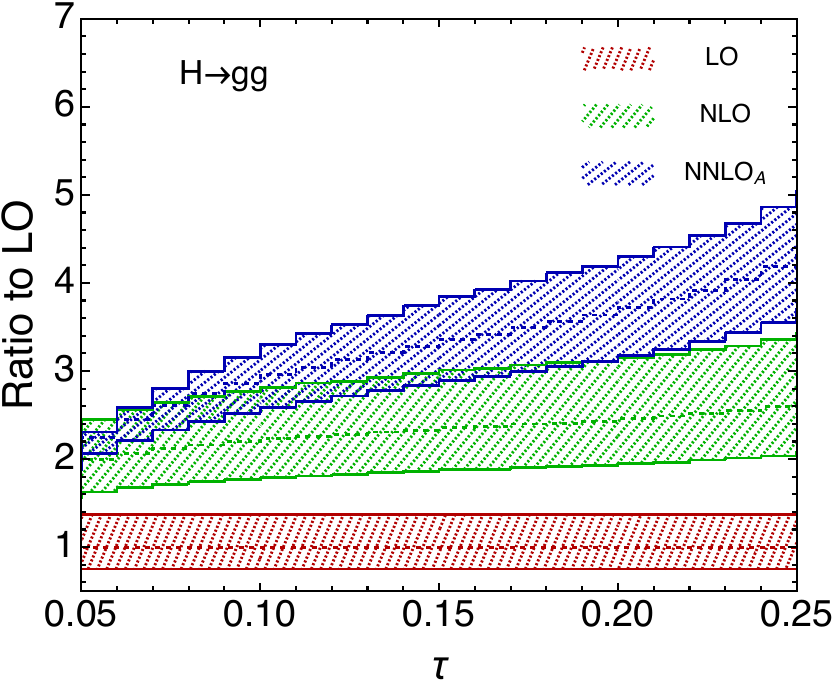}
\includegraphics[width=0.45\textwidth]{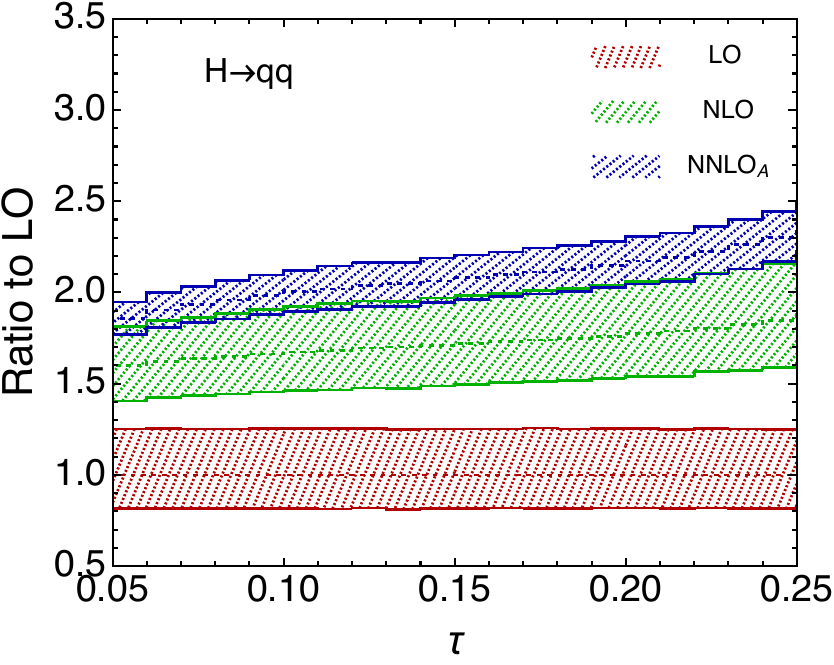}
\caption{\label{fig:NNLOA}Thrust distributions at LO, NLO and approximate NNLO.  }
\end{figure}

In fig.~\ref{fig:NNLOA}, we show the approximate NNLO results for the $Hgg$ and $Hq\bar{q}$ channels in the region $0.05 \leq \tau \leq 0.25$. In the upper plots we show the absolute distributions, while in the lower plots we show the ratios of the differential cross sections to the LO central values. We see that the NNLO corrections are still quite large. Especially for the $Hgg$ channel, the NNLO correction can reach about 50\% of the NLO differential cross section. Nevertheless, the NNLO band now marginally overlaps with the NLO one, indicating that the perturbative series starts to converge. We can therefore expect that the scale variations of the NNLO results provide a relatively honest estimate of the perturbative uncertainties due to missing higher order corrections. 

\begin{figure}[t!]
\centering
\includegraphics[width=0.45\textwidth]{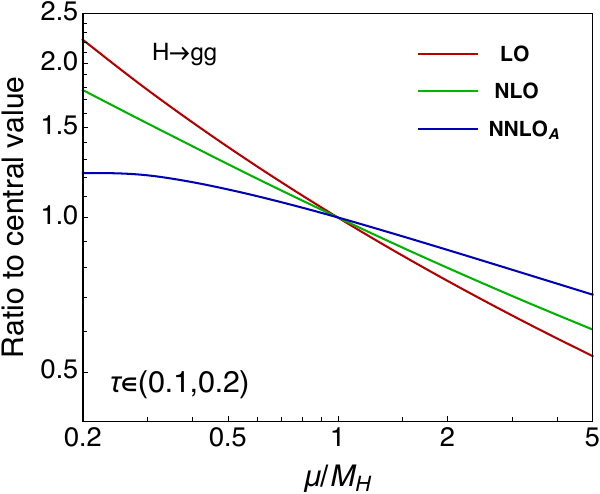}
\quad
\includegraphics[width=0.45\textwidth]{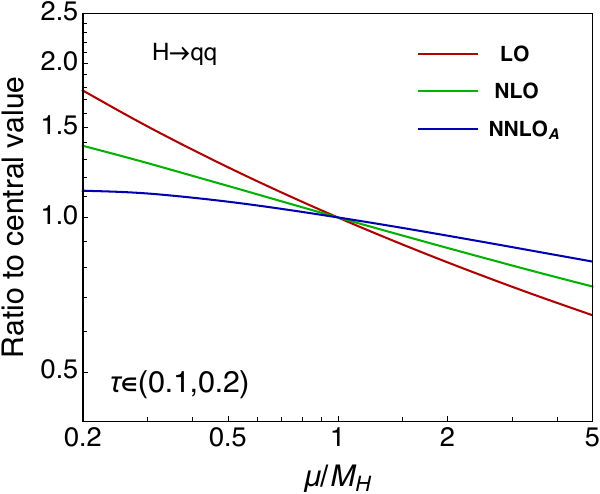}
\caption{\label{fig:scaledep}The ratios of the integrated cross sections in the bin $\tau \in [0.1,0.2]$ to their central values at $\mu=m_H$, as a function of $\mu/m_H$.}
\end{figure}

To see more clearly the relative scale variations at each order, we show in fig.~\ref{fig:scaledep} the ratios of the integrated cross sections in the bin $\tau \in [0.1,0.2]$ to their central values at $\mu=m_H$. The slopes of the curves indicate how strong the predictions depend on the unphysical renormalization scale $\mu$. We observe that the scale dependence consistently decreases as we go to higher orders in perturbation theory. However, for the $Hgg$ channel, the variation of the cross section is still at the level of $\pm 10\%$ when $\mu$ is varied in the range $[m_H/2,2m_H]$, which calls for further improvement to match the precision of future $e^+e^-$ colliders.

\begin{figure}[t!]
  \centering
  \includegraphics[width=0.45\textwidth]{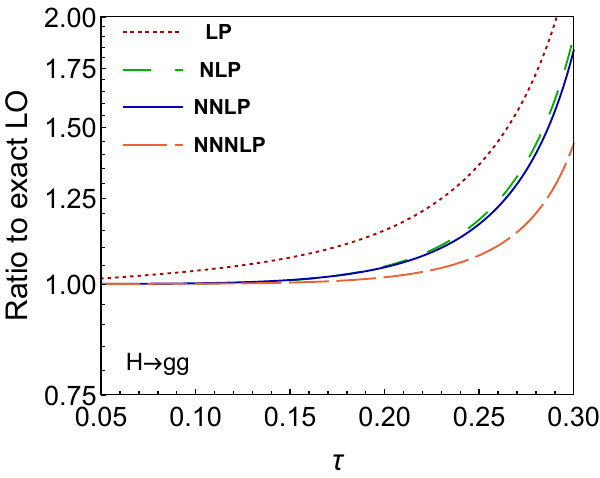}
  \quad
  \includegraphics[width=0.45\textwidth]{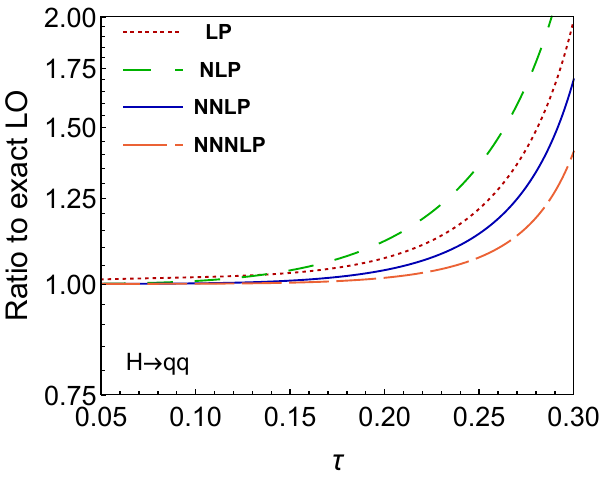}
  \caption{\label{fig:LONLP}The ratios of the LP, NLP, next-to-next-to-leading power (NNLP), and next-to-next-to-next-to-leading power (NNNLP) results to the exact LO.}
\end{figure}

Finally, it should be noted that the factorization formula \eqref{eq:fac}, and hence the leading singular term in Eq.~\eqref{eq:singular}, captures only the leading power (LP) contribution enhanced by $1/\tau$. Recently, there have been a lot of efforts to calculate the next-to-leading power (NLP) corrections for various processes. In particular for thrust distribution, this has been considered in \cite{Moult:2018jjd}. It will be interesting to include such higher power contributions in the approximate NNLO formula. This will improve the accuracy of the approximate formula for moderate $\tau$, and will also extend its range of validity to larger values of $\tau$. While this is beyond the scope of the current work, it is straightforward to perform a power expansion in $\tau$ for the LO distribution using the analytical expressions \eqref{eq:LO}. For example, in the $Hgg$ channel, the result is given by
\begin{align}
  \frac{1}{\Gamma^g_0} \frac{d\Gamma_{\text{LO}}^g}{d\tau} &= \frac{1}{\Gamma^g_0} \frac{d\Gamma_{\text{LO,sing}}^g}{d\tau} \nonumber
  \\
  &+ \frac{\alpha_s^2(\mu)}{\alpha_s^2(m_H)} \frac{\alpha_s(\mu)}{2\pi} \, \Big[ \underbrace{C_A \, \big( 4\ln\tau + 11 \big) - T_F n_f \, \big( 4 \ln\tau + 14 \big)}_{\text{next-to-leading power}} + \mathcal{O}(\tau) \Big] \, .
\label{eq:LOsing}
\end{align}
In Fig.~\ref{fig:LONLP}, we study the convergence of the power expansion for the LO distributions with the central scale choice $\mu=m_H$. We show the ratios of the first 4 orders in the power expansion to the exact LO result. It can be seen that in the $Hgg$ channel, the NLP contribution brings the approximate result much closer to the exact one. On the other hand, in the $Hq\bar{q}$ channel, the NLP result accidentally behaves worses than the LP one for $\tau > 0.15$. Only by including even higher power corrections can one obtain a reliable approximation to the exact LO result. It would be interesting to see in the future whether the same conclusions can be drawn for the NLO and NNLO results.

\section{Conclusion and Outlook}
\label{sec:conclusion}

In this paper, we have presented predictions for the thrust distribution in hadronic decays of the Higgs boson to quarks and gluons.
Our calculation is based on a low energy effective theory with $Hgg$ effective coupling and $Hq\bar{q}$ Yukawa couplings by integrating out the top quark.
We have calculated the NLO QCD corrections to both channels and find large impacts on the differential cross sections. Especially for the di-gluon case, the NLO corrections can be as large as the LO results (corresponding to a $K$-factor $\sim 200\%$).
The scale variations of the LO fail to predict the genuine perturbative uncertainties, and are barely reduced by the inclusion of the NLO corrections.
Besides, the NLO calculation provides a new leading contribution to the large
$\tau$ region $1/3<\tau<(1-1/\sqrt{3})$, in which the LO distribution vanishes.

The above observations indicate that higher order corrections beyond NLO are needed to reduce the perturbative uncertainties of theoretical predictions, in order to match the experimental precision at a future Higgs factory.
As a first step, we have derived an approximate formula based on a factorization
theorem valid in the small $\tau$ limit.
The formula captures the leading singular terms arising from soft and collinear emissions.
We show that the formula provides a reasonable approximation to the exact result for $\tau$ up to $\sim 0.25$ at LO and NLO.
We then use the formula to give an approximate NNLO prediction for the thrust distribution in the range $\tau \in [0.05,0.25]$. We find that the NNLO corrections are still quite sizable and important. They also reduce the scale uncertainties significantly. Therefore, the NNLO results must be taken into account for future experiments.

A couple of improvements over the results in this work are ongoing. First of all, the fixed-order predictions presented in this work cease to be valid in the region of very small $\tau$. In this region, the singular terms $\ln^n\tau/\tau$ in Eq.~\eqref{eq:singular} are too large at each order in $\alpha_s$, such that the perturbative convergence is spoiled. An all-order resummation of these singular contributions is mandatory to arrive at reliable predictions. The ingredients for such a resummation at the next-to-next-to-next-to-leading logarithmic accuracy are available, and can be readily applied. The second improvement concerns the large $\tau$ region. The approximate NNLO formula obtained in this work is not valid there. An exact NNLO calculation would be necessary to correctly describe the tail of the thrust distribution. These improvements will be presented in our forthcoming articles.

\section*{Acknowledgments}

This work was supported in part by the National Natural Science Foundation of China under Grant No. 11575004 and 11635001.
The work of J.~Gao was sponsored by CEPC theory program and by the National Natural Science Foundation of China under the Grant No. 11875189 and No.11835005.
The work of W.~Ju was supported in part by the China Postdoctoral Science Foundation under Grant No. 2017M610685.
Y.~Gong and W.~Ju would like to thank SJTU for hospitality during their stay.

\appendix

\section{Ingredients relevant for LO and NLO calculations}
\label{app:FOingredients}

The $\beta$-function is defined as
\begin{equation}
\frac{d\alpha_s(\mu)}{d\ln\mu} = \beta(\alpha_s) = -2 \alpha_s \sum_{n=0} \left( \frac{\alpha_s}{4\pi} \right)^{n+1} \beta_{n} \, ,
\end{equation}
where the coefficients are given by \cite{vanRitbergen:1997va}
\begin{align}
\beta_{0} &= \frac{11}{3} C_A - \frac{4}{3} n_f T_F \, , \nonumber
\\
\beta_{1} &= \frac{34}{3} C_A^2 - \frac{20}{3} C_A n_f T_F - 4 C_F n_f T_F \, , \nonumber
\\
\beta_{2} &= \frac{325}{54} n_f^2 - \frac{5033}{18} n_f + \frac{2857}{2} \, , \nonumber
\\
\beta_{3} &= \frac{1093}{729} n_f^3 + n_f^2 \left( \frac{6472\zeta_3}{81} + \frac{50065}{162} \right) + n_f \left(-\frac{6508 \zeta_3}{27} - \frac{1078361}{162} \right) + 3564 \zeta_3 + \frac{149753}{6} \, .
\end{align}
Here the color factors are $C_A = N_c$, $C_F = (N_c^2-1)/(2N_c)$, $T_F = 1/2$ and $n_f=5$ is the number of light quarks. For $\beta_2$ and $\beta_3$ we have substituted $N_c = 3$ to shorten the expressions.

The anomalous dimension of the Yukawa coupling $y_q(\mu)$ is the same as the anomalous dimension of quark masses. It is given by
\begin{equation}
\gamma^y(\alpha_s(\mu)) = -\sum_{n=0} \left( \frac{\alpha_s}{4\pi} \right)^{n+1} \gamma^y_{n} \, ,
\end{equation}
with the coefficients given by \cite{Gehrmann:2014vha}
\begin{align}
\gamma^y_0 &= 6 C_F \, , \nonumber
\\
\gamma^y_1 &= 3 C_F^2 + \frac{97}{3} C_A C_F - \frac{10}{3} C_F n_f \, , \nonumber
\\
\gamma^y_2 &= 129 C_F^3 - \frac{129}{2} C_A C_F^2 + \frac{11413}{54} C_A^2 C_F + \left( 48 \zeta_3 - 46 \right) C_F^2 n_f \nonumber
\\
&\hspace{9em} - \left( \frac{556}{27} + 48 \zeta_3 \right) C_A C_F n_f - \frac{70}{27} C_F n_f^2 \, .
\end{align}

The anomalous dimension of $C_t(m_t,\mu)$ is actually not used in our calculation, since we always evaluate the coefficient at the renormalization scale $\mu$ as in eq.~(\ref{eq:GammaNLO}). We nevertheless give it here \cite{Chetyrkin:2005ia} 
\begin{align}
\gamma^t_0 &= 0 \, , \nonumber
\\
\gamma^t_1 &= \frac{40}{3} C_A n_f T_F - \frac{68}{3} C_A^2 + 8 C_F n_f T_F \, , \nonumber
\\
\gamma^t_2 &= -\frac{650}{27} n_f^2 + \frac{10066}{9} n_f - 5714 \, .
\end{align}

\section{Ingredients relevant for the leading singular terms}
\label{app:LSingredients}

We expand the hard Wilson coefficients $C^{i}_S$ in eq.~(\ref{eq:fac}) as
\begin{equation}
C_S^{i}(m_H,\mu) = 1 + \sum_{n=1}^\infty \left( \frac{\alpha_s(\mu)}{4\pi}\right)^n C_S^{i(n)}(L_H)  \, ,
\end{equation}
where
\begin{equation}
   L_H = \ln \frac{-m_H^2-i\epsilon}{\mu^2} \, .
\end{equation}
The NLO and NNLO coefficients are given by \cite{Gehrmann:2010ue, Gehrmann:2014vha}
\begin{align}
C_S^{g(1)}(L_H) &= C_A \left( \frac{\pi^2}{6} - L_H^2 \right) \, , \nonumber
\\
C_S^{g(2)}(L_H) &= C_A^2 \bigg[ \frac{L_H^4}{2} + \frac{11 L_H^3}{9} + \left( \frac{\pi^2}{6} - \frac{67}{9} \right) L_H^2 + \left( -2\zeta_3 - \frac{11\pi^2}{9} + \frac{80}{27 }\right )L_H \nonumber
\\
&\hspace{4em} + \frac{\pi^4}{72} - \frac{143 \zeta_3}{9} + \frac{67 \pi^2}{36} + \frac{5105}{162} \bigg] + C_F n_f \left( 2 L_H + 8 \zeta_3 - \frac{67}{6} \right) \nonumber
\\
&\quad + C_A n_f \left[ -\frac{2 L_H^3}{9} + \frac{10 L_H^2}{9} + \left(\frac{52}{27} + \frac{2 \pi^2}{9} \right) L_H - \frac{46 \zeta_3}{9} - \frac{5\pi^2}{18} - \frac{916}{81} \right] ,
\end{align} 
and
\begin{align}
C_S^{q(1)}(L_H) &= C_F \left( -L_H^2 + \frac{\pi^2}{6} - 2 \right) , \nonumber
\\
C_S^{q(2)}(L_H) &= C_F^2 \left[\frac{L_H^4}{2}+\left(2-\frac{\pi ^2}{6}\right) L_H^2
+ \left( 24 \zeta_3- 2\pi^2 \right) L_H + 6 + \frac{7 \pi ^2}{3} - 30\zeta_3 - \frac{83\pi^4}{360} \right] \nonumber
\\
&+ C_FC_A \left[ \frac{11 L_H^3 }{9} + \left( \frac{\pi^2}{3} - \frac{67}{9} \right) L_H^2 + \left( \frac{242}{27} + \frac{11\pi^2}{9} - 26\zeta_3 \right) L_H \right. \nonumber
\\
& \left. + \frac{151\zeta_3}{9} + \frac{11\pi^4}{45} - \frac{467}{81} - \frac{103\pi^2}{108} \right] + C_Fn_f \left[ -\frac{2 L_H^3}{9} + \frac{10 L_H^2}{9} \right. \nonumber
\\
& \left. - \left( \frac{56}{27} + \frac{2\pi^2}{9} \right) L_H + \frac{2\zeta_3}{9} + \frac{5\pi^2}{54} + \frac{200}{81} \right] .
\end{align}

We now turn to the jet function $J^{i}(s,\mu)$ in eq.~(\ref{eq:fac}). In practice, it is more convenient to work with its Laplace transform
\begin{equation}
\tilde{j}^{i}(L_J,\mu) = \int^{\infty}_{0} ds \, \exp \! \left( -\frac{\nu s}{m_H^2} \right) J^i(s,\mu).
\end{equation}
where
\begin{equation}
L_J = \ln \frac{m_H^2}{\mu^2 \nu e^{\gamma_E}} \, ,
\end{equation}
with $\gamma_E$ the Euler constant. The transformed jet function can be expanded as
\begin{equation}
\tilde{j}^i(L_J,\mu) = 1 + \sum_{n=1}^\infty \left( \frac{\alpha_s(\mu)}{4\pi}\right)^n \tilde{j}^{i(n)}(L_J).
\end{equation}
For our purpose, we need the NLO and NNLO coefficients, as well as the $L_J$-dependent part of the N$^3$LO coefficients. They are given by \cite{Becher:2008cf, Becher:2009th, Bruser:2018rad, Banerjee:2018ozf}
\begin{align}
\tilde{j}^{q(1)}(L_J) &= C_F\left( 2 L_J^2-3 L_J-\frac{2 \pi ^2}{3}+7\right), \nonumber
\\
\tilde{j}^{q(2)}(L_J) &= C_Fn_f \left[ \frac{4}{9} L_J^3 - \frac{29}{9} L_J^2 + \left( \frac{247}{27} - \frac{2\pi^2}{9} \right) L_J + \frac{13\pi^2}{18} - \frac{4057}{324} \right]
\nonumber
\\
&\hspace{-3em} + C_FC_A \left[ -\frac{22}{9} L_J^3 + \left( \frac{367}{18} - \frac{2\pi^2}{3} \right) L_J^2 + \left( 40 \zeta_3 + \frac{11\pi^2}{9} - \frac{3155}{54} \right) L_J - 18 \zeta_3 - \frac{37\pi^4}{180} \right. \nonumber
\\
&\hspace{-3em} \left. -\frac{155\pi^2}{36} + \frac{53129}{648} \right] + C_F^2 \left[ 2 L_J^4 - 6 L_J^3 + \left( \frac{37}{2} - \frac{4\pi^2}{3} \right) L_J^2 + \left( 4\pi^2 - 24 \zeta_3 - \frac{45}{2} \right) L_J
\right.
\nonumber
\\
&\hspace{-3em} \left. - 6 \zeta_3 + \frac{61\pi^4}{90} - \frac{97\pi^2}{12} + \frac{205}{8} \right] \, ,
\nonumber
\\
\tilde{j}^{q(3)}(L_J) &= C_F n_f^2 \Bigg[ \frac{4}{27} L_J^4 - \frac{116}{81} L_J^3 + \left( \frac{470}{81} - \frac{4\pi^2}{27} \right) L_J^2 + \bigg( \frac{58\pi^2}{81} - \frac{8714}{729} - \frac{64}{27} \zeta_3 \bigg) L_J \Bigg]
\nonumber
\\
&\hspace{-3em} + C_FC_An_f \Bigg[ -\frac{44}{27} L_J^4 + \left( \frac{1552}{81} - \frac{8\pi^2}{27} \right) L_J^3 + \left( \frac{28\pi^2}{9} - \frac{7531}{81} + 8 \zeta_3 \right) L_J^2 +\bigg( \frac{32\pi^4}{135}  
\nonumber
\\
&\hspace{-3em} - \frac{1976\zeta_3}{27} - \frac{2632\pi^2}{243} + \frac{160906}{729} \bigg) L_J \Bigg]
+ C_FC_A^2 \Bigg[  \frac{121}{27} L_J^4 + \bigg( \frac{44\pi^2}{27} - \frac{4649}{81} \bigg) L_J^3+ \bigg( \frac{22\pi^4}{45} 
\nonumber
\\
&\hspace{-3em} -132\zeta_3 - \frac{389\pi^2}{27} + \frac{50689}{162} \bigg) L_J^2 + \bigg( \frac{18179\pi^2}{486} - \frac{53\pi^4}{135} - \frac{599375}{729} - 232 \zeta_5 - \frac{88\pi^2\zeta_3}{9}
\nonumber
\\
&\hspace{-3em} + \frac{6688 \zeta_3}{9} \bigg) L_J \Bigg] + C_F^2 n_f \Bigg[ \frac{8}{9} L_J^5 - \frac{70}{9} L_J^4 + \bigg( \frac{875}{27} - \frac{20\pi^2}{27} \bigg) L_J^3 + \bigg( \frac{151\pi^2}{27} - \frac{15775}{162} \bigg) L_J^2
\nonumber
\\
&\hspace{-3em} + \bigg(\frac{32\zeta_3}{9} + \frac{4\pi^4}{27} - \frac{2833\pi^2}{162} + \frac{7325}{36} \bigg) L_J \Bigg]
+ C_F^2C_A \Bigg[ -\frac{44}{9} L_J^5 + \bigg( \frac{433}{9} - \frac{4\pi^2}{3} \bigg) L_J^4
\nonumber
\\
&\hspace{-3em} + \bigg( \frac{164\pi^2}{27} - \frac{10537}{54} + 80 \zeta_3 \bigg) L_J^3 + \bigg( -68 \zeta_3 + \frac{\pi^4}{30} - \frac{2045\pi^2}{54} + \frac{157943}{324} \bigg) L_J^2
\nonumber
\\
&\hspace{-3em} + \bigg( \frac{290\zeta_3}{3} - 120 \zeta_5 - \frac{88\pi^2\zeta_3}{3} - \frac{923\pi^4}{540} + \frac{35075\pi^2}{324} - \frac{151405}{216} \bigg) L_J \Bigg] + C_F^3 \Bigg[  \frac{4}{3} L_J^6 - 6 L_J^5
\nonumber
\\
&\hspace{-3em} + \bigg( 23 - \frac{4\pi^2}{3} \bigg) L_J^4 + \bigg( 8\pi^2 - \frac{99}{2} - 48\zeta_3 \bigg) L_J^3 + \bigg( 60 \zeta_3 + \frac{61\pi^4}{45} - \frac{151\pi^2}{6} + \frac{349}{4} \bigg) L_J^2
\nonumber
\\
&\hspace{-3em} + \bigg( 240 \zeta_5 + \frac{64\pi^2\zeta_3}{3} - 218 \zeta_3 - \frac{149\pi^4}{30} + \frac{145\pi^2}{4} - \frac{815}{8} \bigg) L_J\Bigg] + c_{3q}^J \, ,
\end{align}
and
\begin{align}
\tilde{j}^{g(1)}(L_J) &= C_A \left( 2L_J^2 - \frac{11}{3} L_J + \frac{67}{9} - \frac{2\pi^2}{3}\right) + n_f \left( \frac{2}{3} L_J - \frac{10}{9} \right) , \nonumber
\\
\tilde{j}^{g(2)}(L_J) &= n_f^2 \left( \frac{4}{9} L_J^2 - \frac{40}{27} L_J - \frac{2\pi^2}{27} + \frac{100}{81} \right)
+ C_Fn_f \left( 2L_J + 8\zeta_3 - \frac{55}{6} \right) + C_An_f \Bigg[ \frac{16}{9} L_J^3
\nonumber
\\
&\hspace{-3em} - \frac{28}{3} L_J^2 + \bigg( \frac{224}{9} - \frac{10\pi^2}{9} \bigg) L_J - \frac{8\zeta_3}{3} + \frac{67\pi^2}{27} - \frac{760}{27} \Bigg]
+ C_A^2 \Bigg[ 2 L_J^4 - \frac{88}{9} L_J^3 + \bigg( \frac{389}{9} - 2\pi^2 \bigg) L_J^2
\nonumber
\\
&\hspace{-3em} + \bigg( \frac{55\pi^2}{9} + 16\zeta_3 - \frac{2570}{27} \bigg) L_J - \frac{88\zeta_3}{3} + \frac{17\pi^4}{36} - \frac{362\pi^2}{27} + \frac{20215}{162} \Bigg] \, ,
\nonumber
\\
\tilde{j}^{g(3)}(L_J) &= n_f^3 \Bigg[ \frac{8}{27} L_J^3 - \frac{40}{27} L_J^2 + \bigg( \frac{200}{81} - \frac{4\pi^2}{27} \bigg) L_J \Bigg] + C_F n_f^2 \Bigg[ \frac{10}{3} L_J^2 + \bigg( 16\zeta_3 - 24 \bigg) L_J \Bigg]
\nonumber
\\
&\hspace{-3em} - C_F^2n_f L_J + C_An_f^2 \Bigg[ \frac{4}{3} L_J^4 - \frac{292}{27} L_J^3 + \bigg( \frac{3326}{81} - \frac{4\pi^2}{3} \bigg) L_J^2 + \bigg( \frac{508\pi^2}{81} - \frac{116509}{1458} - \frac{256\zeta_3}{27} \bigg) L_J \Bigg]
\nonumber
\\
&\hspace{-3em} + C_AC_Fn_f \Bigg[ \frac{16}{3} L_J^3 + \big( 32\zeta_3 - 55 \big) L_J^2 + \bigg( -\frac{8\pi^4}{45} - \frac{10\pi^2}{3} + \frac{5599}{27} - \frac{1096\zeta_3}{9} \bigg) L_J \Bigg]
\nonumber
\\
&\hspace{-3em} + C_A^2n_f \Bigg[ \frac{20}{9} L_J^5 - \frac{64}{3} L_J^4 - \bigg( \frac{88\pi^2}{27} - \frac{3106}{27} \bigg) L_J^3 + \bigg( \frac{586\pi^2}{27} - \frac{8\zeta_3}{3}     -\frac{10067}{27} \bigg)  L_J^2
\nonumber
\\
&\hspace{-3em} + \bigg( \frac{449\pi^4}{270} - \frac{16831\pi^2}{243} + \frac{1052135}{1458} - \frac{1280\zeta_3}{27} \bigg) L_J \Bigg] + C_A^3 \Bigg[ \frac{4}{3} L_J^6 - \frac{110}{9} 
L_J^5 + \bigg( 85 - \frac{8\pi^2}{3} \bigg) L_J^4
\nonumber
\\
&\hspace{-3em} + \bigg( \frac{484\pi^2}{27} - \frac{9623}{27} + 32 \zeta_3 \bigg) L_J^3 + \bigg( \frac{169\pi^4}{90} - \frac{484\zeta_3}{3} - \frac{2362\pi^2}{27} + \frac{85924}{81} \bigg) L_J^2
\nonumber
\\
&\hspace{-3em} + \bigg( -\frac{4411\pi^4}{540} + \frac{52678\pi^2}{243} - \frac{1448021}{729} - 112\zeta_5 - \frac{160\pi^2\zeta_3}{9} + \frac{6316\zeta_3}{9} \bigg) L_J \Bigg] + c^J_{3g} \, .
\end{align}
The $L_J$-independent terms $c^J_{3q}$ and $c^J_{3g}$ are known, but are not relevant to the calculations in this work.

The case for the soft function $S^i(k,\mu)$ is similar. We define its Laplace transform as
\begin{equation}
\tilde{s}^i(L_S,\mu) = \int_0^\infty dk \, \exp \! \left( -\frac{\nu k}{m_H} \right) S^i(k,\mu) \, ,
\end{equation}
where
\begin{equation}
L_S = \ln \frac{m_H}{\mu \nu e^{\gamma_E}} \, .
\end{equation}
Again, we need the expansion coefficients of $\tilde{s}^i(L_S,\mu)$ up to the NNLO and the $L_S$-dependent terms at N$^3$LO. They can be written as \cite{Becher:2008cf, Kelley:2011ng} 
\begin{align}
\tilde{s}^{q(1)}(L_S) &= C_F \left( -8L_S^2 - \pi^2 \right) ,
\nonumber
\\
\tilde{s}^{q(2)}(L_S) &= C_Fn_f \Bigg[ -\frac{32}{9} L_S^3 + \frac{80}{9} L_S^2 - \bigg( \frac{8\pi^2}{9} + \frac{224}{27} \bigg) L_S - \frac{52\zeta_3}{9} + \frac{77\pi^2}{27} + \frac{40}{81} \Bigg]
\nonumber
\\
&\hspace{-3em} + C_F C_A \Bigg[ \frac{176}{9} L_S^3 + \bigg( \frac{8\pi^2}{3} - \frac{536}{9} \bigg) L_S^2 + \bigg( \frac{44\pi^2}{9} - 56 \zeta_3 + \frac{1616}{27} \bigg) L_S + \frac{286\zeta_3}{9} + \frac{14\pi^4}{15}
\nonumber 
\\
&\hspace{-3em} - \frac{871\pi^2}{54} - \frac{2140}{81} \Bigg]
+ C_F^2 \Bigg( 32 L_S^4 + 8\pi^2 L_S^2 + \frac{\pi^4}{2}\Bigg) \, ,
\nonumber
\\
\tilde{s}^{q(3)}(L_S) &= C_Fn_f^2 \Bigg[ -\frac{64}{27} L_S^4 + \frac{640}{81} L_S^3 - \bigg (\frac{32\pi^2}{27} + \frac{800}{81} \bigg) L_S^2 + \bigg( \frac{64\pi^2}{9} - \frac{3200}{729} - \frac{64\zeta_3}{9} \bigg) L_S \Bigg]
\nonumber
\\
&\hspace{-3em} + C_FC_An_f \Bigg[ \frac{704}{27} L_S^4 + \bigg( \frac{64\pi^2}{27} - \frac{9248}{81} \bigg) L_S^3 + \bigg( \frac{64\pi^2}{9} + \frac{16408}{81} \bigg) L_S^2 + \bigg( \frac{6032\zeta_3}{27} + \frac{64\pi^4}{45}
\nonumber 
\\
&\hspace{-3em} - \frac{19408\pi^2}{243} - \frac{80324}{729} \bigg) L_S \Bigg]
+ C_FC_A^2 \Bigg[ - \frac{1936}{27} L_S^4 - \bigg( \frac{352\pi^2}{27} - \frac{28480}{81} \bigg) L_S^3 + \bigg( \frac{104\pi^2}{27} 
\nonumber
\\
&\hspace{-3em} - \frac{88\pi^4}{45} - \frac{62012}{81} + 352 \zeta_3  \bigg) L_S^2 + \bigg( \frac{50344\pi^2}{243} - \frac{88\pi^4}{9} + \frac{556042}{729} + 384\zeta_5 + \frac{176\pi^2\zeta_3}{9} 
\nonumber
\\
&\hspace{-3em} - \frac{36272\zeta_3}{27} \bigg) L_S \Bigg]
+ C_F^2n_f \Bigg[ \frac{256}{9} L_S^5 - \frac{640}{9} L_S^4 + \bigg( \frac{32\pi^2}{3} + \frac{1504}{27} \bigg) L_S^3 + \bigg( \frac{5620}{81} - \frac{856\pi^2}{27}
\nonumber
\\
&\hspace{-3em} - \frac{160\zeta_3}{9} \bigg) L_S^2 + \bigg( \frac{608\zeta_3}{9} + \frac{56\pi^4}{45} + \frac{152\pi^2}{27} - \frac{3422}{27} \bigg) L_S \Bigg]+ C_F^2C_A \Bigg[ - \frac{1408}{9} L_S^5
\nonumber
\\
&\hspace{-3em} + \Bigg( \frac{4288}{9} - \frac{64\pi^2}{3} \Bigg) L_S^4 + \bigg( 448 \zeta_3 - \frac{176\pi^2}{3} - \frac{12928}{27} \bigg) L_S^3 + \bigg( \frac{5092\pi^2}{27} - \frac{2288\zeta_3}{9} - \frac{152\pi^4}{15}
\nonumber
\\
&\hspace{-3em} + \frac{17120}{81} \bigg) L_S^2
+ \bigg( 56 \pi ^2 \zeta_3 - \frac{44\pi^4}{9} - \frac{1616\pi^2}{27} \bigg) L_S \Bigg] + C_F^3 \Bigg[ -\frac{256}{3} L_S^6 - 32 \pi^2 L_S^4 - 4 \pi^4 L_S^2 \Bigg]
\nonumber
\\
&\hspace{-3em} + c_{3q}^S \, ,
\end{align}
where again the constant term $c_{3q}^S$ is not relevant for this work. The expression for the gluon soft function can be obtained from the quark one by a Casimir scaling.

\section{Leading singular terms up to NNLO}
\label{app:singular}

In Eq.~\eqref{eq:singular}, the singular parts of thrust distributions are expressed in terms of the coefficients $\Delta_{i}^{(n)}(\tau,\mu)$. Here we give their explicit expressions, where we set the number of colors $N_c = 3$ for simplicity. We also set $\mu = m_H$ to get rid of the scale-dependent logarithms, and one can easily recover them through the RG equation. For the $Hgg$ channel, the results are given by
\begin{align}
\Delta_g^{(1)}(\tau,m_H) &=
\left(\frac{4 n_f}{3}-22\right) \frac{1}{\tau} - 24 \frac{\ln(\tau)}{\tau} \, ,
\\
\Delta_g^{(2)}(\tau,m_H) &=
\left[ 360 \zeta_3 - 88\pi^2 - 2150 + \left( \frac{16\pi^2}{3} + \frac{640}{3} \right) n_f - \frac{40}{9} n_f^2 \right] \frac{1}{\tau} \nonumber
\\
&\hspace{-3em} + \left( \frac{8n_f^2}{3} - 8n_f - 120\pi^2 - 1410 \right) \frac{\ln(\tau)}{\tau} + \left( 1188 - 72 n_f \right) \frac{\ln^2(\tau)}{\tau} + 288 \frac{\ln^3(\tau)}{\tau} \, ,
\\
\Delta_g^{(3)}(\tau,m_H) &=
\bigg[ \left( \frac{256}{9} n_f^2 - 368 n_f - 1672 \right) L_{HT} + \left( \frac{800}{81} - \frac{80\pi^2}{81} \right) n_f^3 \nonumber
\\
&\hspace{-3em} + \left( \frac{1304\pi^2}{27} - \frac{992\zeta_3}{3} - \frac{31081}{27} \right) n_f^2 + \left( \frac{742121}{27} - \frac{4276\pi^2}{9} + 7552\zeta_3 - \frac{26\pi^4}{15} \right) n_f \nonumber
\\
&\hspace{-3em} - 37152\zeta_5 + 3456\pi^2\zeta_3 - 20904\zeta_3 + \frac{143\pi^4}{5} - \frac{698\pi^2}{3} - \frac{1610351}{9} \bigg] \frac{1}{\tau} \nonumber
\\
&\hspace{-3em} + \bigg[ - \left( 512 n_f + 1824 \right) L_{HT} - \frac{320}{27} n_f^3 + \left( \frac{352\pi^2}{9} + \frac{5512}{9} \right) n_f^2 \nonumber
\\
&\hspace{-3em} + \left( 7072\zeta_3 - 896\pi^2 - \frac{2044}{3} \right) n_f - 90288 \zeta_3 - \frac{372\pi^4}{5} - 568 \pi^2 -\frac{205012}{3} \bigg]  \frac{\ln(\tau)}{\tau} \nonumber
\\
&\hspace{-3em} + \bigg[ \frac{32}{9} n_f^3 + 144 n_f^2 - \left( 624 \pi^2 + 11616 \right) n_f - 26784 \zeta_3 + 10296 \pi^2 + 126876 \bigg] \frac{\ln^2(\tau)}{\tau} \nonumber
\\
&\hspace{-3em} + \bigg[ -\frac{1184}{9} n_f^2 + \frac{9184}{3} n_f + 2304\pi^2 - 3752 \bigg] \frac{\ln^3(\tau)}{\tau} + \left(960 n_f-15840\right)  \frac{\ln^4(\tau)}{\tau} \nonumber
\\
&\hspace{-3em} - 1728 \frac{\ln^5(\tau)}{\tau} \, ,
\label{eq:exact expressions singular Hgg}
\end{align}
where $L_{HT}=\ln(m_H/m_t)$. For the $Hq\bar{q}$ process, we have
\begin{align}
\Delta_q^{(1)}(\tau,m_H) &= - \frac{8}{\tau}  - \frac{32}{3} \frac{\ln(\tau)}{\tau} \, ,
\\
\Delta_q^{(2)}(\tau,m_H) &=
\left( \frac{40}{3} n_f + \frac{1120\zeta_3}{9} - \frac{128\pi^2}{9} - 340 \right) \frac{1}{\tau} + \left(
\frac{176}{27} n_f - \frac{160\pi^2}{9} - \frac{2056}{9} \right) \frac{\ln(\tau)}{\tau} \nonumber
\\
&\hspace{1em} + \left( 304 - \frac{32}{3} n_f \right) \frac{\ln^2(\tau)}{\tau} + \frac{512}{9} \frac{\ln^3(\tau)}{\tau} \, ,
\\
\Delta_q^{(3)}(\tau,m_H) &= 
\Bigg[ \left( \frac{1952\pi^2}{243} - \frac{1024}{27} \zeta_3 - \frac{3056 }{81} \right) n_f^2 + \left( \frac{106624 \zeta_3}{81} + \frac{608 \pi^4}{405} - \frac{6880 \pi^2}{27} \right. \nonumber
\\
&\hspace{-4em} \left. + \frac{16640}{9} \right) n_f - \frac{42688 \zeta_5}{9}+\frac{8192 \pi ^2 \zeta_3}{27}-\frac{198016 \zeta_3 }{27}-\frac{10472 \pi ^4}{405} + \frac{48248 \pi^2}{27} - \frac{516776}{27} \Bigg]  \frac{1}{\tau} \nonumber
\\
&\hspace{-4em} + \Bigg[ \left( \frac{128\pi^2}{81} + \frac{1120}{243} \right) n_f^2 + \left( \frac{19840\zeta_3}{27} - \frac{22400\pi^2}{243} + \frac{31376}{81} \right) n_f - \frac{118208 \zeta_3 }{9} \nonumber
\\
&\hspace{-4em} - \frac{2336\pi^4}{1215} + \frac{76064\pi^2}{81} - \frac{193688}{27} \Bigg] \frac{\ln(\tau)}{\tau} + \Bigg[ \frac{544}{27} n_f^2 - \left( \frac{1216\pi^2}{27} + \frac{4160}{3} \right) n_f -\frac{86528 \zeta_3}{27} \nonumber
\\
&\hspace{-4em} + \frac{9632\pi^2}{9} + \frac{53576}{3} \Bigg] \frac{\ln^2(\tau)}{\tau} + \left( -\frac{896}{81} n_f^2 + \frac{33152}{81} n_f + \frac{11264\pi^2}{81} - \frac{9632}{3} \right) \frac{\ln^3(\tau)}{\tau} \nonumber
\\
&\hspace{-4em} + \left( \frac{2560}{27} n_f - \frac{6400}{3} \right) \frac{\ln^4(\tau)}{\tau} - \frac{4096}{27} \frac{\ln^5(\tau)}{\tau} \, .
\end{align}

\end{document}